\documentclass[aps,prl,twocolumn,showpacs,superscriptaddress,groupedaddress]{revtex4}

\usepackage{graphicx}  
\usepackage{dcolumn}  
\usepackage{bm}       
\usepackage{bbm}
\usepackage{amssymb}
\usepackage{amsmath}
\usepackage{tikz}
\usetikzlibrary{decorations.shapes}
\usetikzlibrary{arrows}
\usetikzlibrary{patterns}

\begin{document}

\widetext

\title{Solution to sign problems in models of interacting fermions and quantum spins}
\author{Emilie Huffman}
\affiliation{Department of Physics, Box 90305, Duke University,
Durham, North Carolina 27708, USA}
\author{Shailesh Chandrasekharan}
\affiliation{Department of Physics, Box 90305, Duke University,
Durham, North Carolina 27708, USA}
\affiliation{Center for High Energy Physics, Indian Institute of Science, Bangalore, 560 012, India}

\date{\today}

\begin{abstract}
We show that solutions to fermion sign problems in the CT-INT formulation, where the path integral is expanded in powers of the interaction in continuous time, can be extended to systems involving fermions interacting with dynamical quantum spins. While these sign problems seem unsolvable in the auxiliary field approach, solutions emerge in the worldline representation of quantum spins. Combining this idea with meron-cluster methods, we are able to further extend the class of models that are solvable. We demonstrate these novel solutions to sign problems by considering several examples of strongly correlated systems that contain the physics of semimetals, insulators, superfluidity, and antiferromagnetism.
\end{abstract}
\maketitle

\section{Introduction}
Simulating quantum many body systems using path integral Monte Carlo methods--even for systems in equilibrium--remains challenging due to the sign problem, which can even be NP-hard in some cases \cite{Tro05}. While this means that finding a general solution applicable to all systems is unlikely, particular solutions can still be found that are applicable to specific systems. Solutions found so far fall under three broad categories: (1) Finding the right basis for the Hilbert space such that sign problems are absent. The most recent example is the solution to a class of frustrated quantum spin systems \cite{Alet16alm}. (2) Finding a resummation of the partition function that renders the resummed weights positive. This includes methods like the meron-cluster method \cite{Bietenholz:1995zk,Chandrasekharan:1999cm}, the subset method \cite{Alford:2001ug,Bloch:2011jx} and the fermion bag approach \cite{Cha10,Cha12,Chandrasekharan:2013rpa}. (3) Finding a symmetry such that every term of the sum can be written as a square of real number. Solutions to fermion sign problems, especially for systems in more than one spatial dimension, combine ideas from (2) and (3) \cite{Blankenbecler:1981jt,Hands:2000ei,Fehske08,Drut:2012md,Lee:2008fa}. While most solutions to sign problems so far have been obtained in equilibrium systems, some sign problems in systems experiencing purely dissipative dynamics in real time have also been solved for strongly interacting quantum spin systems \cite{PhysRevB.90.241104,PhysRevB.92.035116,PhysRevB.92.121104}. In addition to sampling configurations that arise in the path integral, another promising approach is to directly sample Feynman diagrams that arise in perturbation theory using a Monte Carlo method. Here the sign problem has a different origin and progress can be made within a class of problem \cite{Pro98,PhysRevLett.99.250201,PhysRevB.77.125101}. Recently this approach has been applied to solve a class of sign problems in real time \cite{PhysRevLett.115.266802}. 

In this work we only focus on the path integral formulation in imaginary time. Even in such cases many sign problems especially in systems containing fermions, remain unsolvable due to sign problems. Recently it was discovered that, when the fermionic path integral can be expanded in powers of the interaction in continuous time (the CT-INT formalism), fermion sign problems can be solved in certain cases \cite{Rub05,Bon06,Bur08,Gou10,Gul11}. The fermion bag approach is an extension of this idea to discrete time formulations and to strong couplings \cite{Chandrasekharan:2011vy,PhysRevD.85.091502}. Recently, we applied the idea to solve the sign problem in a class of spin-polarized systems by exploiting particle-hole symmetry \cite{Huf14}. Our solution was later formulated in the Majorana representation, which makes the pairing mechanism between particles and holes more explicit and can even be used to construct an auxiliary field approach to the problem \cite{Li15}.  Additional guiding principles involving the concepts of Majorana reflection positivity and Time Reversal symmetry have also been found, and help extending the solvable class of models \cite{Wei16,Li16}. 

Quantum Monte Carlo (QMC) techniques in the CT-INT formalism for fermionic models are currently being developed \cite{PhysRevB.91.241118}. While a naive quantum Monte Carlo method for the CT-INT formulation scales as $\beta^3 N^3$ (where $\beta$ is the extent of the imaginary time and $N$ is the number of spatial lattice sites), recently it was recast in a more efficient form that allows one to construct Monte Carlo methods that scale like $\beta N^3$ (the LCT-QMC approach) \cite{Wan15}, which is similar to the auxiliary field approach. Some simulations have also been performed in order to compute the critical exponents at the quantum phase transition, using both the LCT-QMC and the auxiliary field methods \cite{Wan14,Li15a}. In addition, the CT-INT formalism is very similar to the fermion bag approaches developed for lattice field theories \cite{Cha10,Cha12,Chandrasekharan:2013rpa}. It has been recently shown that one can perform calculations on large lattices by cleverly storing information necessary to perform quick updates within large space-time regions \cite{PhysRevD.93.081701}. We believe these ideas could easily be extended to solve all problems that can be formulated in the CT-INT formulation. Interestingly, bosonic models can also be formulated in the CT-INT formulation through the stochastic series expansion and updated efficiently with either the directed loop algorithms \cite{PhysRevE.66.046701,PhysRevE.71.036706} or worm updates \cite{PhysRevLett.87.160601}. Thus, it is clear that once problems involving either bosons or fermions can be formulated without a sign problem in the CT-INT formulation one can readily construct QMC methods to solve them numerically.

In this paper we extend the class of models solvable through the CT-INT formulation, to include those that contain interactions between fermions and quantum spins. The essential idea behind our current work was introduced in lattice field theory to solve a sign problem in a Yukawa model involving interacting fermions and bosons \cite{Chandrasekharan:2012fk}. In that work the fermion bag approach was used to solve the fermion sign problem, while at the same time the worldline representation was used to solve the bosonic sign problem that arises in the fermion bag approach. This idea to combine the solution of the sign problem in the fermion sector with an appropriate solution to the sign problem in the bosonic sector, can naturally be extended to a variety of models involving fermions and quantum spins interacting with each other. This idea has also been used recently in the impurity Monte Carlo developed recently where the sign problem turns out to be mild \cite{Elhatisari:2014lka}. In this work we also show that the class of solvable models may be further broadened by allowing frustrating interactions in the bosonic sector that can only be solved using the meron cluster idea. While we show the absence of the sign problem in the CT-INT representation, the LCT-QMC representation will also have no sign problem. Hence it should be possible to develop algorithms that scale as $\beta N^3$ for these problems. We must note that systems with interacting fermions and bosons in the Hamiltonian formulation have been solved before, but these systems either had a non-interacting bosonic bath that allowed one to integrate out the bath variables  \cite{Ass14:Pavarini:155829}. Our work extends this idea further.  

In order to demonstrate the new class of solutions we consider several models in this work and show how their partition functions can be written as a sum of positive terms, such that each term can be calculated in polynomial time. In section 2 we introduce our ideas by considering a simple extension of the spin polarized $t-V$ model by coupling it to the transverse field quantum Ising model. In Section 3 we introduce antiferromagnetism by replacing the Ising model with the Heisenberg quantum antiferromagnet. In section 4, we introduce a model that requires the use of the meron cluster idea in the spin sector to solve the sign problem. Section 5 shows how the ideas can be extended to a class of $SU(2)$ symmetric models, including the classic problem of the half filled Kondo-lattice model. Section 6 contains our conclusions.

\section{CT-INT with Bosonic Worldlines}

In this section we introduce our ideas by considering a simple extension of the t-V model that we solved recently \cite{Huf14}, by coupling it to the transverse field quantum spin-half Ising model.We will also develop the notation that will be helpful in later sections. The Hamiltonian of the system we consider is given by
\begin{equation}
    \begin{aligned}
    H = & -\lambda \sum_{\left\langle ij\right\rangle} \left(c_i^\dagger c_j + c^\dagger_j c_i \right) + V \sum_{\left\langle ij\right\rangle}\left(n_i-\frac{1}{2}\right) \left(n_j - \frac{1}{2}\right) \\
    & - J \sum_{\left\langle ij\right\rangle} S_i^z S_j^z   + \sum_i h_i \left(n_i-\frac{1}{2}\right) S_i^x,
    \end{aligned}
    \label{ising}
\end{equation}
where $S_i^a$ are the quantum spin-half operators, $c_i^\dagger$ and $c_i$ are creation and annihilation operators of spinless fermions on the lattice site $i$ of a bipartite lattice, $\langle ij\rangle$ refers to nearest neighbor sites where we assume $i$ and $j$ belong to opposite sublattices. The first term on the right hand side is the free fermion term $H_0^f$ and the third term will be referred to as the free boson term $H_0^b$. The second term, which we refer to as $H_{\rm int}^f$, creates repulsive interactions between nearest neighbor fermions that live on opposite sublattices (i.e., we assume $V \geq 0$). The fourth term, referred to as $H_{\rm int}^{fb}$, couples fermions with bosons and mimics a fluctuating transverse field depending on the fermion occupation on that site. We assume the remaining couplings $\lambda$, $J$ and $h_i$ are real but arbitrary. Although the focus of this work is not to uncover the physics of the above model, we believe it has a rich phase diagram on a honeycomb lattice where two orders compete. In the absence of quantum spins, the fermions can be in a semi-metal or a Mott insulating phase. It is interesting to ask if these phases can coexist with or destroy the Ising order of the quantum Ising model when the two sectors are coupled. A naive reasoning suggests that in the semi-metal phase the $H_{\rm int}^{fb}$ term is expected to be small and the Ising order in the spin sector can survive. However, in the Mott insulating phase the $H_{\rm int}^{fb}$ term is strong and can destroy the Ising order. The phase diagram should also have interesting quantum critical points with gapless fermions.

While we cannot rule out a clever auxiliary field approach to the above problem, at least naively such an approach seems impossible. The reason for this is that to integrate out the fermions one would naturally choose a spin basis that diagonalizes $S_i^x$. But in that basis, the fermion determinants are not positive for all background spin configurations. As we already know from previous work, for a positive determinant one needs a staggered chemical potential \cite{Huf14},
\begin{equation}
H_{\rm stagg} = \sum_i h_i \sigma_i \left(n_i - \frac{1}{2}\right),
\end{equation}
where $\sigma_i$ is the parity of a site (i.e., $+1$ for one sublattice and $-1$ for the other), and $h_i \geq 0$ for all $i$. In the above problem the fluctuating quantum variable $S_i^x$ would destroy this property. The associated sign problem in this example is very similar to the one encountered in \cite{Chandrasekharan:2012fk} and hence the solution is also very similar. In order to solve it, we first transform the Hamiltonian with a unitary transformation
\begin{equation}
H^U = U^\dagger H U,\qquad U = \prod_{i} \mathrm{e}^{i(1-\sigma_i)  S_i^z \pi/2}
\label{utran}
\end{equation}
such that all the terms in $H$ remain unchanged except for the fermion-boson coupling, which is transformed into:
\begin{equation}
H_{\rm int}^{U,fb} = \sum_i h_i \sigma_i \big(n_i - \frac{1}{2}\big) S_i^x
\end{equation} 
In the transformed basis we perform the CT-INT expansion of the partition function \cite{Rub05,Bon06,Bur08,Gou10,Gul11},
\begin{equation}
\begin{aligned}
Z = \sum_l  &\int_0^\beta  ...\int_0^{t_3}\int_0^{t_2} dt_1 dt_2...dt_l \; (-1)^l \\
& \times  {\rm Tr}\left(e^{-\left(\beta - t_{1}\right) H_0} H_{\rm int} e^{-\left(t_1 -t_2\right) H_0} H_{\rm int}...\right),
\label{ctint}
\end{aligned}
\end{equation}
where there are $l$ insertions of $H_{\rm int}$ in the trace at times $t_1,...,t_l$, and we take $H_0= H_0^f + H_0^b$ and $H_{\rm int} = H_{\rm int}^f + H_{\rm int}^{U,fb}$. From now on we use the symbol $\left[dt\right]$ as a shorthand for all such time-ordered integrals. In the expansion (\ref{ctint}), we have operators in two different spaces: the fermionic space and the spin space. Since each spin operator commutes with each fermionic operator we can factorize the trace in each term of the expansion into a product of two traces: one trace over the spin states containing only operators in the spin space and one trace over the fermionic states containing operators only in the fermionic space. For example, here is one of the terms in the expansion at order $l=2$ with two insertions of interactions, one insertion of $H_{\rm int}^f$ at $t_1$ and another insertion of $H_{\rm int}^{U,fb}$ at $t_2$:
\begin{equation}
\begin{aligned}
    \left(-1\right)^{2} {\rm Tr}&\Big(e^{-\left(\beta - t_2\right)H_0^b}  h_kS_k^x e^{-t_2 H_0^b}\Big) \\
 &\times {\rm Tr}\Big(e^{-\left(\beta - t_1\right)H_0^f} V\big(n_i-\frac{1}{2}\big)\big(n_j-\frac{1}{2}\big)\\
&\qquad\times e^{-\left(t_1 - t_2\right) H_0^f}  \sigma_k\big(n_k - \frac{1}{2}\big) e^{-t_2  H_0^f}\Big).
    \end{aligned}
\end{equation}
Using this factorization, the partition function can be written as:
\begin{equation}
\begin{aligned}
Z = \sum_{l,\{k\},m,\left\{b\right\}}  \int & \left[dt\right] \\
&\times G_s\left[l,\left\{k\right\}\right] G_f \left[l, \left\{k\right\},m,\left\{b\right\}\right],
\end{aligned}
\label{ctintf}
\end{equation}
where
\begin{equation}
\begin{aligned}
G_s\left[l,\left\{k\right\}\right] & = (-1)^l {\rm Tr}\left(e^{-\left(\beta - t_1\right)H_0^b}  h_{k_1} S_{k_1}^x \right. \\
& \times \left. e^{-\left(t_1 - t_2\right) H_0^b} h_{k_2} S_{k_2}^x .... h_{k_l} S_{k_l}^x e^{-\left(t_{l}\right) H_0^b}\right),
\end{aligned}
\end{equation}
is the trace over the spin space, and depends on insertions of $l$ insertions of the interaction terms $h_kS_k^x$ at the times $t_1,t_2,...,t_l$. Similarly, 
\begin{equation}
\begin{aligned}
& G_f \left[l, \left\{k\right\}, m,\left\{b\right\}\right] = (-1)^m {\rm Tr}\left(...\sigma_{k_1}\left(n_{k_1} -1 / 2\right)...\right.\\
& \quad \left. ...H_{\rm int}^f(b_1) ... H_{\rm int}^f(b_m) ...\sigma_{k_l}\left(n_{k_l} -1 / 2\right) ...\right).
\end{aligned}
\label{ftrace}
\end{equation}
is the trace in the fermionic space and depends on $m$ insertions of the interaction bonds $H_{\rm int}^f(b\equiv \langle ij\rangle) = V\left(n_i - \frac{1}{2}\right) \left(n_j - \frac{1}{2}\right)$ and $l$ insertions of $\sigma_k\left(n_k -\frac{1}{2}\right)$ from the fermion-spin interactions. One such configuration of insertions is labeled by $[l,\{k\},m,\{b\}]$. The presence of the free propagators $\mathrm{e}^{-t H_0^f}$ between these insertions are hidden in the ellipses. Note that for every insertion at of $S_k^x$ at $t_k$ in the spin space, we have a corresponding insertion of $\sigma_k\left(n_k - \frac{1}{2}\right)$ at $t_k$ in the fermionic space. This provides correlations between the two spaces. The partition function is a sum over all possible configurations $[l,\{k\},m,\{b\}]$.

\begin{figure}
\begin{center}
\includegraphics{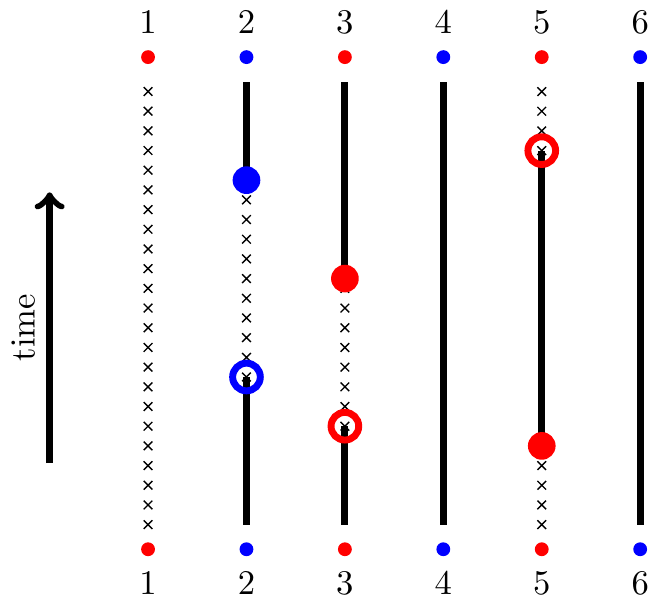}
\end{center}
\caption{An illustration of a hardcore boson worldline configuration in one spatial dimension. The spatial sites are numbered $1-6$, with odd sites colored red and even sites colored blue. Among the wordlines a cross indicates the absence of the boson and a solid line indicates its presence. Insertions of $S_x$ create or annihilate hardcore bosons. Filled circles indicate a creation event while the open circle indicates the annihilation event. On each site the $S_x$ operators come in pairs due to temporal periodicity of the worldlines.}
\label{sx}
\end{figure}

We already know from \cite{Huf14}, the insertion of $\sigma_k\left(n_k -\frac{1}{2}\right)$ along with the factor $(-1)^m$ ensures that the trace in the fermionic space $G_f[l,\{k\},m,\{b\}] \geq 0$. Let us now argue that the trace in the spin space is also positive. We evaluate the trace in the $S^z$ basis by inserting the identity $I = \sum_{s^z} \left|s^z\right\rangle \left\langle s^z \right|$ after every insertion of the $S_k^x$. We get
\begin{equation}
\begin{aligned}
& G_s\left[l,\left\{k\right\}\right] = \sum_{\left\{s^z\left(t\right)\right\}}  \left\langle s^z\left(t_0\right)\right| e^{-\left(\beta-t_1\right) H_0^b} S_{k_1}^x \left|s^z\left(t_1\right)\right\rangle  \\
\times & \left\langle s^z\left(t_1\right)\right| e^{-\left(t_1 -t_2\right) H_0^b} S_{k_2}^x \left|s^z\left(t_2\right)\right\rangle ... \left\langle s^z\left(t_l\right)\right| e^{-t_l H_0^b} \left|s^z\left(t_0\right)\right\rangle .
\end{aligned}
\end{equation}
where the sum over $\left\{s^z\left(t\right)\right\}$ indicates a sum over all space-time spin configurations that are periodic  i.e., $s^z\left(t_0\right) = s^z\left(t_l\right)$. Because $H_0^b$ is diagonal in the chosen basis, propagators $\mathrm{e}^{-t H_0^b}$ are just numbers and do not change the spin configuration. On the other hand $S_i^x = \frac{1}{2}\left(S_i^- + S_i^+\right)$, so an insertion of $S_i^x$ flips the spin at the site $i$. Thus, the spin configurations contain spin flips at space-times points $(k_0,t_0), (k_1,t_1), ... (k_l,t_l)$. However, since the configurations need to be periodic at each spatial site $i$ the number of insertions of $S_i^x$ must come in pairs, although they may come at different times. For this reason $l$ is always even and spin trace only depends on $h_i^2$. Another way to view the above scenario is to consider quantum spins as hardcore bosons (with spin-up representing particles and spin-down representing their absence). Then, for every creation (annihilation) of a particle caused by the $S_i^x$ operator, we require a corresponding annihilation (creation) of the same particle caused by a second $S_i^x$ operator to preserve the trace. Due to this constraint, the spin trace $G_s[l,\{k\}] \geq 0$. In FIG.~\ref{sx} we show a pictorial illustration of an allowed hardcore boson configuration. Since both spin and fermion traces can be evaluated in polynomial time, we conclude that (\ref{ising}) has no sign problem in the CT-INT formulation when quantum spins are formulated in the bosonic worldline representation.

\section{Adding anti-ferromagnetism}

\begin{center}
\begin{figure}
\begin{center}
\includegraphics{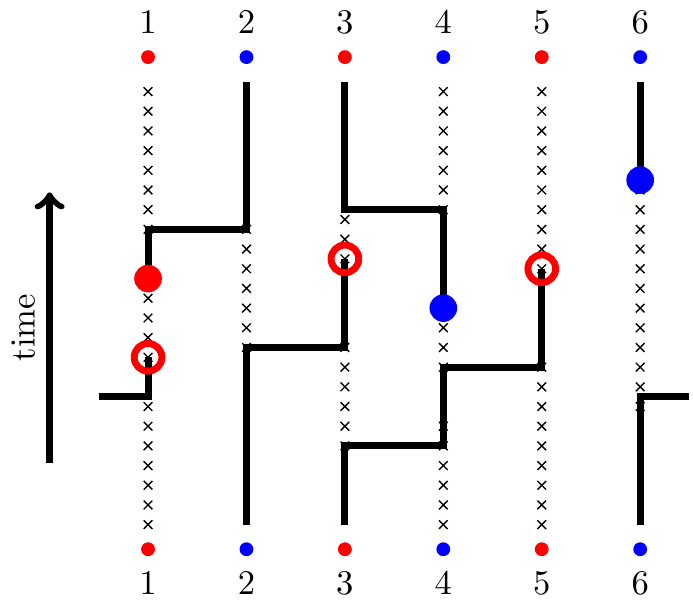}
\end{center}
\caption{Worldline diagram for Heisenburg model. Again, the spatial sites are numbered $1-6$ with red dots for odd sites and blue dots for even sites.}
\label{heisenwl}
\end{figure}
\end{center}

In the model of the previous section, we considered the spin sector to have the simplest possible self interaction, namely the Ising interaction. Clearly it would be interesting to replace this with the full $SU(2)$ symmetric antiferromagnetic interaction. While the Ising interaction forced the $S_i^x$ to come in pairs on each site in the partition function, in the presence of antiferromagnetism this condition is no longer necessary. Although the $S_i^x$ terms do still come in pairs, they need not be on the same site. Despite this complication, the sign problem is solvable for a class of models as we show below. To see this consider the model where $H_0^b$ is replaced with the Heisenberg antiferromagnet. The Hamiltonian is now given by
\begin{equation}
    \begin{aligned}
    H = & -\lambda\sum_{\left\langle ij\right\rangle} \left(c_i^\dagger c_j + c^\dagger_j c_i \right) + V\sum_{\left\langle ij\right\rangle}\left(n_i-\frac{1}{2}\right) \left(n_j - \frac{1}{2}\right) \\
    &+ J\sum_{\left\langle ij\right\rangle} \vec{S}_i \cdot \vec{S}_j + \sum_i h_i \left(n_i-\frac{1}{2}\right) S_i^x.
    \end{aligned}
    \label{heisen}
\end{equation}
For antiferromagnetism, we now require $J\geq 0$. We then need to set $h_i\geq 0$ for all $i$ (or equivalently $h_i\leq 0$ for all $i$) for the solution to the sign problem. To proceed we first perform the unitary transformation (\ref{utran}) as before. The Heisenberg term transforms to $H_0^{b} + H_{\rm int}^{U,b}$, where
\begin{equation}
\begin{aligned}
H_0^{b} & =J\sum_{\left\langle ij \right\rangle} S_i^z S_j^z \\
H_{\rm int}^{U,b} &=  - \frac{J}{2}
\sum_{\left\langle ij \right\rangle} \left( S_i^+ S_j^- + S_i^- S_j^+\right).
\end{aligned}
\label{spinsplit}
\end{equation}
In addition, the fermion-spin interaction is transformed as before to $H_{\rm int}^{U,fb}$. Now the interaction consists of three terms $H_{\rm int} = H_{\rm int}^f + H_{\rm int}^{U,b}+H_{\rm int}^{U,fb}$. Expanding the partition function as in the previous section, we obtain an expression similar to $(\ref{ctintf})$:
\begin{equation}
\begin{aligned}
Z = \sum_{m,\left\{b\right\}} \sum_{n,\left\{h\right\}} \sum_{l,\left\{k\right\}} & \int \left[dt\right] 
\quad G_s\left[n,\left\{d\right\},l,\{k\}\right]  \\
& \times G_f\left[l,\left\{k\right\}, m, \left\{b\right\}\right],
\end{aligned}
\label{pfmod2}
\end{equation}
where the spin trace is given by
\begin{eqnarray}
&&G_s\left[n,\left\{d\right\},l,\{k\}\right] \ =\ (-1)^{l+n}{\rm Tr}\Big(... h_{k_1}S_{k_1}^x... \nonumber \\
&& \qquad ...H^{U,b}_{\rm int}(d_1)...h_{k_2}S_{k_2}^x ... H^{U,b}_{\rm int}(d_n) ... h_{k_l}S_{k_l}^x ...\Big).
\label{gsmod2}
\end{eqnarray}
Now the trace depends on $n$ insertions of nearest neighbor spin hops $H_{\rm int}^{U,b}(d\equiv \langle ij\rangle) = -(J/2) \left( S_i^+ S_j^- + S_i^- S_j^+\right)$, and as before $l$ insertions of $h_kS_k^x$ with the free propagator $\mathrm{e}^{-t H_0^b}$ in between represented as ellipses. This configuration is labeled with $[n,\{d\},l,\{k\}]$. The fermionic trace is the same as before and is given by (\ref{ftrace}), where each configuration is labeled by $[l,\{k\},m,\{b\}]$. As in the previous example it is positive. The bosonic trace is also positive since $l$ turns out to be even and the $(-1)^n$ factor is cancelled by the negative signs that appear in front of $H_{\rm int}^{U,b}(d)$. The trace in the spin space is evaluated by inserting a complete set of states in the $S^z$ basis as before. Each insertion of $S^x_i$ flips a single spin on the site $i$, while the insertion of $H_{\rm int}^{U,b}(d)$ flips both spins on the bond denoted by $d$. In the language of hard core bosons, $S^x_i$ acts as either a creation or an annihilation event while $H_{\rm int}^{U,b}(d)$ acts as an event where the boson hops. Since every creation event needs to be accompanied by an annihilation event, $l$ must be even as previously stated but not necessarily on the same site. An illustration of the hardcore boson configuration is shown  in FIG.~\ref{heisenwl}. Thus, again there is no sign problem in the CT-INT expansion when spins are represented in the worldline representation. While we have focused on a model containing anti-ferromagnetism in this section, it is easy to extend our results to models containing superfluidity.

\section{Extensions with meron clusters}

The two models that we considered above had the property that with an appropriate unitary transformation, the CT-INT approach combined with a worldline formulation for spins naturally led to positive weights. However, many interesting models do not fall in this class and the weights of the configurations in the CT-INT approach continue to be negative. It would be interesting to find methods to solve such remnant sign problems. We would like to argue that in a subset of these models, the solution to the sign problem can be obtained via a resummation over the spin configurations. Thus, the remnant sign problem in these models is hidden in the spin sector and not in the fermion sector. We illustrate this through an example, where the required resummation is performed using the meron cluster idea.

Consider the model we studied in the previous section, but with a slightly modified fermion-spin coupling, $H_{\rm int}^{fb}$. The Hamiltonian is given by
\begin{equation}
\begin{aligned}
H = & -\lambda\sum_{\left\langle i,j \right\rangle} \left(c^\dagger_i c_j + c^\dagger_j c_i\right) + V\sum_{\left\langle i,j \right\rangle} \left(n_i - \frac{1}{2}\right) \left(n_j - \frac{1}{2}\right) \\
& +J \sum_{\left\langle i,j \right\rangle} \vec{S}_i \cdot \vec{S}_j - h\sum_i \sigma_i \left(S_i^x + \frac{1}{2}\right) \left(n_i - \frac{1}{2}\right).
\end{aligned}
\end{equation}
Unlike in Eq.~(\ref{heisen}), there is already a $\sigma_i$ factor in the $H_{\rm int}^{fb}$ term, and instead of $S_i^x$ there is $S_i^x + 1/2$. Further we use $-h$ instead of the general $h_i$ and assume $h$ is positive. Due to the presence of $\sigma_i$, it is better not to perform the unitary transformation since the fermionic trace needs that factor for positivity. Then proceeding as before, it is tempting to split the fermion-spin coupling into two interaction terms:
\begin{equation}
\sigma_i \left(S_i^x + \frac{1}{2}\right) \left(n_i - \frac{1}{2}\right)
= \sigma_i S_i^x \left(n_i - \frac{1}{2}\right) + \frac{\sigma_i}{2} \left(n_i - \frac{1}{2}\right)
\label{fssplit}
\end{equation}
and treat them as separate interactions in the CT-INT expansion. However, such a treatment leads to sign problems. To see this let us proceed as in the previous example except that we treat the second term on the right hand side of (\ref{fssplit}) as a new interaction that appears in the fermionic sector. Since in the CT-INT expansion the interaction terms of this new form are similar to other interactions that already appear within the fermionic trace, $G_f$ continues to be positive, as expected. On the other hand, the spin trace is given by the same equation as (\ref{gsmod2}) but with $H_{\rm int}^{U,b}\left(d\right)$ insertions replaced by $-H_{\rm int}^{U,b}\left(d\right)$ insertions (since we did not perform the unitary transformation) and all the $h_k$ factors replaced with $-h$ factors. Hence, the factor $(-1)^n$ in the front no longer cancels with the negative sign in front of $H_{\rm int}^{U,b}$ as in the previous example. Performing the unitary transformation would only push the problem into the fermionic sector by removing the necessary $\sigma_i$ factors from the fermionic interactions. Also notice that the sign of a configuration depends on $n$ and not the details of the spin configuration. Thus, any resummation over the spin configurations would not help.

The solution is to treat the left hand side of (\ref{fssplit}) as one piece and perform the full trace over the spin space. As we will argue below this can indeed be accomplished in polynomial time. To see this, let us first modify the Heisenberg anti-ferromagnetic term by adding an irrelevant constant to it and treating the whole term as an interaction:
\begin{equation}
H_{\rm int}^b \ =\  - J \sum_{\left\langle i,j \right\rangle} \ \left(\frac{1}{4}-\vec{S}_i \cdot \vec{S}_j\right).
\end{equation}
Since in this new approach every term containing the quantum spin variable is treated as an interaction, we set $H_0^b = 0$. The partition function is identical to (\ref{pfmod2}), where the fermionic trace is the same as before and is given by (\ref{ftrace}), which is clearly positive. On the other hand, the spin trace is different and is given by
\begin{eqnarray}
&& G_s\left[n,\left\{d\right\},l,\{k\}\right] =(-1)^{l+n}{\rm Tr}\Big( ... (-h(1/2+S_{k_1}^x))...
\nonumber \\
&& \ \  ...H^b_{\rm int}(d_1) ... H^b_{\rm int}(d_n) ... (-h (1/2+S_{k_l}^x)) ...\Big).
\label{gsmod3}
\end{eqnarray}
The trace depends on the $n$ insertions of nearest neighbor spin interaction $H_{\rm int}^b(d\equiv \langle ij\rangle) = -J(1/4-\vec{S}_i \cdot \vec{S}_j)$, and $l$ insertions of $-h (1/2+S_k^x)$ with no free propagators between these insertions. The configuration is labeled with $[n,\{d\},l,\{k\}]$. Clearly the $(-1)^{n+l}$ in the front on the right hand side of Eq.~(\ref{gsmod3}) cancels the negative factors in the interaction terms.

\begin{figure}[h]
\includegraphics[width=0.4\textwidth]{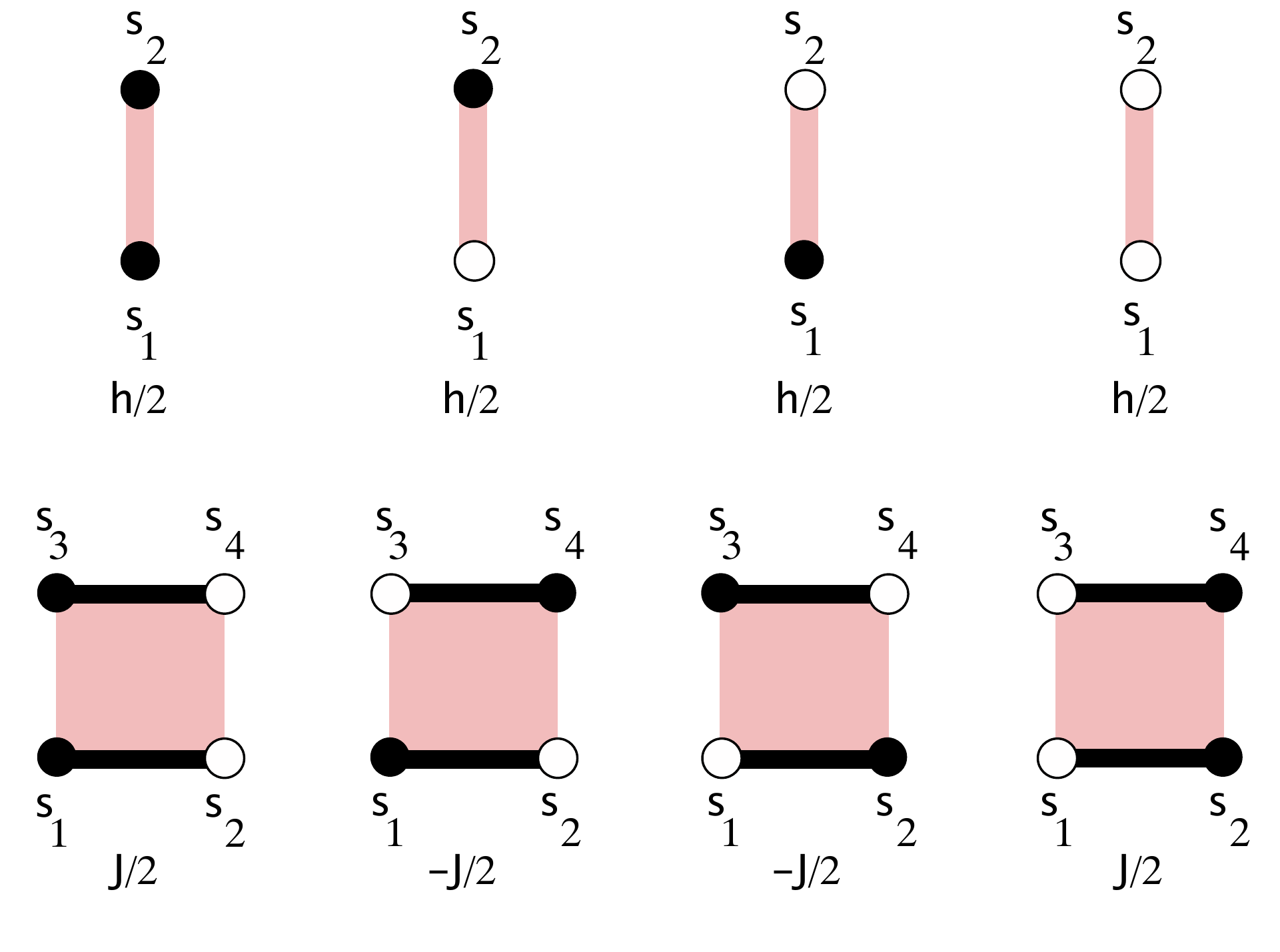}
\caption{Top row: Four non-zero matrix elements that result from an insertion of $(1/2+S^x)$ on a site. All four have the same weight $h/2$. The filled circle represents spin-up and empty circle represents spin-down. Bottom row: Four non-zero matrix elements that result from an insertion of $(1/4-\vec{S}_i \cdot \vec{S}_j)$ on a bond connecting neighboring sites. All four have weights with the same magnitude $J/2$. The off diagonal terms have negative signs.}
\label{matelem}
\end{figure}

In order to compute the spin trace we introduce a complete set of eigenstates of the $S^z$ operator at each site between the interactions to obtain
\begin{equation}
\begin{aligned}
G_s&\left[n,\left\{d\right\},l,\left\{k\right\}\right] =(-1)^{n+l}\\ 
\times&\sum_{\left\{s^z\left(t\right)\right\}}  \left\langle s^z\left(t_0\right)\right| ...\left(-h\left(1/2+ S_{k_1}^x\right)\right) \left|s^z\left(t_{k_1}\right)\right\rangle \left\langle s^z\left(t_{k_1}\right)\right|  \\
&\qquad\qquad ...H^b_{\rm int}\left(d_1\right)\left|s^z\left(t_{d_1}\right)\right\rangle \left\langle s^z\left(t_{d_1}\right)\right|...H^b_{\rm int}\left(d_n\right)\\
& \qquad\qquad\times\left|s^z\left(t_{d_n}\right)\right\rangle\left\langle s^z\left(t_{d_n}\right)\right|...\left(-h\left(1/2+ S_{k_l}^x\right)\right)\\
&\qquad\qquad\times\left|s^z\left(t_{d_l}\right)\right\rangle \left\langle s^z\left(t_{d_l}\right)\right|...\left|s^z\left(t_{0}\right)\right\rangle .
\end{aligned}
\end{equation}
In the $S^z$ basis we know that the interaction $(1/2+S_k^x)$ is a $2 \times 2$ matrix on the single site $k$ and $(1/4-\vec{S}_i \cdot \vec{S}_j)$ is a $4\times 4$ matrix on the bond $d = \langle ij\rangle$. The matrix elements of these interaction matrices can be viewed as providing correlations among the spin degrees of freedom that are involved in the interaction and given a diagrammatic representation. For example, in the $S^z$ basis $s = (\uparrow,\downarrow)$ we find
\begin{equation}
\Big\langle s_1 \Big|
h\Big(\frac{1}{2} + S^x\Big)\Big|s_2\Big\rangle \ =\ \frac{h}{2},
\end{equation}
i.e., all four matrix elements are equal to $h/2$. In FIG.~\ref{matelem} these matrix elements are shown as diagrams that contain two disconnected circles representing the spins $s_1$ and $s_2$. The fact that there is no line connecting the two spins refers to the fact that the two spin degrees of freedom $s_1$ and $s_2$ are completely uncorrelated and independent of each other. Also since each configuration has the same weight $h/2$ each spin can be flipped without affecting the weight of the configuration. Similarly, the non-zero matrix elements of 
\begin{equation} 
\Big\langle s_3 s_4\Big| J \Big(\frac{1}{4}-\vec{S}_i \cdot \vec{S}_j\Big)\Big| s_1 s_2\Big\rangle \ =\ \ \frac{J}{2} (\tau_2)_{s_4s_3}\ (\tau_2)_{s_1s_2},
\end{equation}
where $\tau_2$ is the second Pauli matrix, are also shown in FIG.~\ref{matelem}. These figures show two sets of anti-correlated spins $s_1 s_2$ and $s_3 s_4$. We represent the anti-correlations with a horizontal bond. This means if  $s_1=\uparrow$, then $s_2=\downarrow$, and vice versa.  As long as these anti-correlations are maintained, the non-zero matrix elements have the same magnitude. However, in this case when the spin pair flips (or spins exchange), the weight of the diagram (or the matrix element) is negative. In other words, if $s_1=s_3$ and $s_2=s_4$, i.e. the spins do not flip, the matrix element is positive, but when $s_1 = s_4$ and $s_2 = s_3$, i.e. the spin flips, the matrix element is negative.

\begin{figure}[h]
\includegraphics[width=0.35\textwidth]{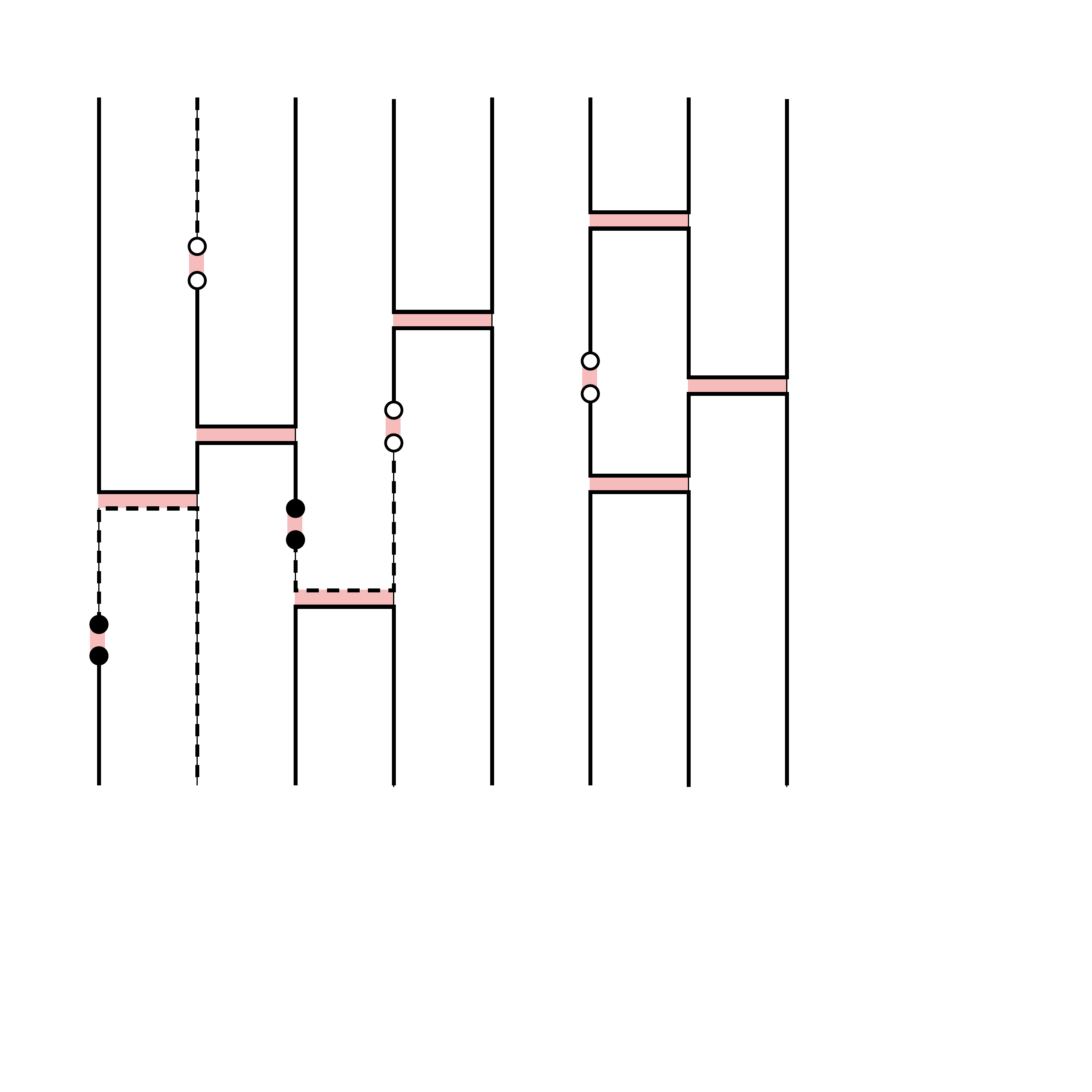}
\caption{An illustration of a cluster configuration that emerges uniquely from a given configuration $\left[n,\left\{d\right\},l,\{k\}\right]$ of operator insertions. Open lines that end on two different sublattices turn out to be a meron cluster, shown as dashed lines in the figure. The spin trace $G_s$ vanishes if the cluster configuration contains a meron.}
\label{clustconf}
\end{figure}

In the calculation of $G_s\left[n,\left\{d\right\},l,\{k\}\right]$ we multiply the interaction matrices in a time ordered pattern. Diagrammatically, we can arrange them at appropriate space-times locations and multiply them such that the second spin label of the previous matrix matches the first spin label on the later matrix on the same lattice site. Thus,  $G_s\left[n,\left\{d\right\},l,\{k\}\right]$ is nothing but a tensor network, which can be pictorially viewed as a network of vertical straight lines connecting identical spins on different matrices that are arranged in a time ordered pattern. Combining this picture with the information in FIG.~\ref{matelem}, that the interaction matrix elements themselves provide correlations among spins, every configuration $\left[n,\left\{d\right\},l,\{k\}\right]$, can be mapped uniquely to a collection of open lines and closed loops in space-time. Each open line or a closed loop is referred to as a cluster. An illustration of this cluster configuration is shown in FIG.~\ref{clustconf}.

Each spin configuration can still be negative. However, the sum over all spin configurations (i.e., the spin trace $G_s\left[n,\left\{d\right\},l,\{k\}\right]$) turns out to be either positive or zero. When performing a trace, if two spins are correlated (or anti-correlated) they must be counted as a single spin. Thus, every cluster should be treated as a correlated object and visualized as a single spin degree of freedom that can exist in two different states. Therefore, if there are $N_C$ clusters in the configuration $\left[n,\left\{d\right\},l,\{k\}\right]$, the computation of the trace requires one to a sum over $2^{N_c}$ spin configurations. Can we find a way to compute this sum over an exponentially large number of terms quickly?

Interestingly, this sum was already performed in an earlier study using the meron cluster approach \cite{Chandrasekharan:2001ya}. While the earlier work used a discrete time method to formulate the trace, it is possible to work directly in continuous time as well \cite{Bea96}. The idea is to first note that a flip of the spins within a cluster can only potentially change the sign of the configuration but not its magnitude, because as long as correlations of the spins are maintained, the magnitude of the matrix elements do not change. Further, negative signs arise only from clusters that contain spin hops. If a cluster is an open line whose end points lie on different sub-lattices, then the line has an odd number of hops. Hence its flip will change the total number of spin exchanges from odd (even) to even (odd) and change the sign of a configuration. Such open lines are referred to as meron clusters. This property of the cluster does not depend on the state of spins of other clusters. Hence, the bosonic trace vanishes in the presence of a meron cluster. Only cluster configurations without any meron clusters make non-zero contributions to the trace. Interestingly, we can always flip all clusters such that one sub-lattice contains up spins and the opposite sub-lattice contains down spins. This is referred to as a reference configuration, and it always has a positive weight. Hence, in a cluster configuration with no merons, all cluster flips come with positive sign and add up. Defining $N_m$ as the number of meron clusters and $N_c$ as the total number of clusters, the meron cluster approach shows that
\begin{equation}
G_s\left[n,\left\{d\right\},l,\{k\}\right] = (J/2)^n \ (h/2)^l\  2^{N_c}\ \delta_{N_m,0},
\end{equation}
which is positive and easily computable. 

The quantum Hamiltonian presented in this section illustrates that meron cluster methods for spin systems may be combined with the CT-INT approach to extend the class of solvable sign problems in combined Bose-Fermi systems.

\section{SU(2) Symmetric Models}
In the three examples we considered so far, we neglected the fermion spin. In this section we illustrate an example of how we can also include spin and continue to work in the CT-INT formulation in a class of $SU(2)$ symmetric models interacting with quantum spins. In the model we consider, quantum spins interact with fermions through an $SU(2)$ symmetric interaction. The Hamiltonian is given by
\begin{eqnarray}
H &=& -\lambda\sum_{\left\langle ij \right\rangle, \sigma} \left(c^\dagger_{i,\sigma} c_{j,\sigma} + c^\dagger_{j,\sigma} c_{i,\sigma}\right) + h \sum_i \vec{S}_i \cdot c_i^\dagger \vec{\tau} c_i 
\nonumber \\
&&\qquad \ + \  J\ \sum_{\langle ij\rangle} \vec{S}_i\cdot\vec{S}_j,
\label{su2mod}
\end{eqnarray}
where now the fermion creation and annihilation operators also carry the spin index $\sigma=\uparrow,\downarrow$, $\vec{\tau}$ are Pauli matrices in this space, and in the second term on the right hand side we have used the spinor notation $c_i^\dagger \equiv \left(\begin{array}{cc} c^\dagger_{i,\uparrow} & c_{i,\downarrow}^\dagger\end{array}\right)$. We can rewrite the interaction term between spins and fermions as
\begin{equation}
h\sum_{i} S_i^+ c_{i,\downarrow}^\dagger c_{i,\uparrow} + h\sum_{i} S_i^- c_{i,\uparrow}^\dagger
 c_{i,\downarrow} +h \sum_{i} S_i^z \left(n_{i,\uparrow} - n_{i,\downarrow}\right),
\end{equation}
Using the transformations $c_{i,\downarrow}\rightarrow\sigma_i c^\dagger_{i,\downarrow}$, $c_{i,\downarrow}^\dagger\rightarrow\sigma_i c_{i,\downarrow}$, and Eq.~(\ref{utran}) we obtain the following transformed Hamiltonian (up to an overall constant):
\small \begin{equation}
\begin{aligned}
H = &-\lambda\sum_{\left\langle ij \right\rangle, \sigma} \left(c^\dagger_{i,\sigma} c_{j,\sigma}+ c^\dagger_{j,\sigma} c_{i,\sigma}\right) + h \sum_{i,\sigma} S_i^z \ n_{i,\sigma} \\
& +h \sum_{i} S_i^+ c_{i,\downarrow} c_{i,\uparrow} + h \sum_{i} S_i^- c_{i,\uparrow}^\dagger c_{i,\downarrow}^\dagger \\
&  - J\ \sum_{\left\langle ij \right\rangle} \Big(\frac{1}{4}  - S_i^z S_j^z \Big)  - \frac{J}{2} \sum_{\left\langle ij \right\rangle} \left( S_i^+ S_j^- + S_i^- S_j^+\right).
\end{aligned}
\label{sumodel}
\end{equation}
We treat all terms on the right hand side of the above equation in the first line as the free fermionic Hamiltonian $H_0^f$. The two terms in the second line are treated as two different fermion-spin couplings $H_{\rm int}^{fs,a},a=1,2$. The terms in the last line are treated as $H_{\rm int}^b$ as in the previous section. Performing the usual CT-INT expansion we obtain,
\begin{equation}
\begin{aligned}
Z = & \sum_{l^1,l^2,m}  \int  \left[dt\right] \ (-1)^{l^{(1)}+l^{(2)}+m} 
\times  {\rm Tr}\Big(.... H_{\rm int}^b(d_1)... \\
&  ... H_{\rm int}^{fs,1}(k^1_1)
 ... H_{\rm int}^{fs,2}(k^2_1) ... H_{\rm int}^b(d_m)... H_{\rm int}^{fs,2}(k^2_{l^2})...\Big),
\label{ctintfinal}
\end{aligned}
\end{equation}
where the ellipses stand for the free fermion propagators. In the trace we have $l^{(1)}$ insertions $H_0^{fs,1}(k)=h S_k^+ c_{k,\downarrow} c_{k,\uparrow}$ and $l^{(2)}$ insertions of
$H_0^{fs,2}(k)=hS_i^- c_{k,\uparrow}^\dagger c_{k,\downarrow}^\dagger$ and $m$ insertions of the bond operator $H_{\rm int}^b(d=\langle ij\rangle) = - J \Big(1/4 -S_i^z S_j^z\Big)  - J/2\left( S_i^+ S_j^- + S_i^- S_j^+\right)$.  Due to spin and fermion number conservation, it is clear we must have $l^{(1)} = l^{(2)}$. Further the $(-1)^m$ cancels the negative signs that comes from $m$ insertions of $H_{\rm int}^b(d)$. Unfortunately, now the trace cannot be factored into a product of a trace over the fermion space and a trace over the spin space. However if we evaluate the spin trace in the $S^z$ basis then, as we have explained in the previous section, insertions of $H_{\rm int}^b$ can be mapped uniquely into a cluster configuration of correlated spins. If there are $N_c$ clusters in the configuration, the full spin trace for a fixed insertions of $H_{\rm int}^b$ is a sum over $2^{N_c}$ spin flips. Although this sum cannot be performed explicitly as in the previous example, the weight of each of the $2^{N_c}$ spin configurations can be computed and the partition function can be written as
\begin{equation}
\begin{aligned}
Z = \sum_{l^1,l^2,m}  \int & \left[dt\right] \  (J/2)^m(h)^{l^{(1)}+l^{(2)}}
\sum_{[s_i(t)]} {\rm Tr}_f\Big(.... c_{k^1_1,\downarrow}c_{k^1_1,\uparrow} \\
& ... c_{k^2_1,\uparrow}^\dagger c_{k^2_1,\downarrow}^\dagger
 ...  c_{k^2_{l^{(2)}},\uparrow}^\dagger c_{k^2_{l^{(2)}},\downarrow}^\dagger... c_{k^1_{l^{(1)}},\uparrow}c_{k^1_{l^{(1)}},\downarrow}...\Big),
\end{aligned}
\end{equation}
where the spin trace appears as a sum over $2^{N_c}$ spin configurations represented as $[s_i(t)]$, and the fermion trace still appears in the expression. Unlike previous examples, it depends on the background spin configuration $[s_i(t)]$ through the free propagators that appear in the ellipses. Since the fermion spins do not mix with each other and appear symmetrically, the fermion trace factors into two identical terms, one is a trace over the spin up space and the other over the spin down space. Each of these can be expressed as a determinant of a matrix $M[s_i(t)]$ that depends on the spin configuration. The exact expression for $M[s_i(t)]$ can be obtained using the usual Wick's theorem \cite{Negele:1988vy}. Thus, we finally obtain the expression
\begin{equation}
Z = \sum_{l^1,l^2,m}  \sum_{[s_i(t)]} \int  \left[dt\right] \ (J/2)^m (h)^{l^{(1)}+l^{(2)}}  \Big(\mathrm{Det}\big(M[s_i(t)]\big)\Big)^2.
\end{equation}
Thus, there is no sign problem the CT-INT expansion. A simple reduction of the above model gives the well known Kondo-lattice model at half filling, whose Hamiltonian is given by
\begin{eqnarray}
H &=& -\lambda\sum_{\left\langle ij \right\rangle, \sigma} \left(c^\dagger_{i,\sigma} c_{j,\sigma} + c^\dagger_{j,\sigma} c_{i,\sigma}\right) + h \sum_{i\in {\cal L}} \vec{S}_i \cdot c_i^\dagger \vec{\tau} c_i.
\label{klat}
\end{eqnarray}
In this model, fermions interact with a lattice of spin impurities located at the sites $i\in{\cal L}$. It can be obtained from Eq.(\ref{su2mod}) by setting $J=0$ and assuming that spins are located only at a subset of lattice sites. While the Kondo-lattice problem at half filling is also solvable with the usual auxiliary field Monte Carlo method \cite{PhysRevLett.83.796}, we believe that an alternate approach such as the one presented here is useful, since it helps to view the problem in different light. Of course a solution to the more difficult sign problem away from half filling, where the Kondo lattice model is considered as the microscopic model for heavy fermion systems \cite{PhysRevLett.57.877}, would be truly exciting.

\section{Conclusions}
In this work we have shown that for a class of systems consisting of fermions interacting with quantum spins (or hardcore bosons), the CT-INT approach leads to new representations of the partition function that do not suffer from the sign problem. In addition to fermions interacting with spins, both fermions and spins can in principle interact with themselves thus allowing one to solve a rich variety of systems with Monte Carlo calculations. While we considered only four specific examples in this article in order to explain our ideas concretely, a careful reader will recognize that our methods extend to many more problems, especially ones that include disorder. We have also argued that the solvable class may be expanded further by combining our ideas with the meron-cluster technique in the bosonic sector. The fact that cluster algorithms and techniques for quantum spin models can be married naturally with the CT-INT approach is exciting.

It would be interesting to understand if the class of problems we are able to solve with our ideas naturally fall within some framework like Majorana reflection positivity, proposed in \cite{Wei16}. It is likely that such a framework exists if one can view quantum spins in analogy with fermions. We have not focused on Monte Carlo methods or the efficiency of the CT-INT expansion for these problems. More work is perhaps needed to ensure efficient calculations.

We thank Lei Wang for explaining to us the details of the LCT-QMC method and other helpful discussions. We also would like to thank Uwe-Jens Wiese for many discussions and collaboration over the years. We thank ECT* at Trento for hosting a conference on Diagrammatic Monte Carlo methods in Nuclear, Particle and Condensed Matter Physics, that motivated us to work on this project. SC would like to thank the Center for High Energy Physics at the Indian Institute of Science for hospitality, where part of this work was done. The material presented here is based upon work supported by the U.S. Department of Energy, Office of Science, Nuclear Physics program under Award Number DE-FG02-05ER41368. EH is also supported by a National Physical Science Consortium fellowship.

\bibliography{ref,refsch}

\end{document}